\documentclass[fleqn,10pt]{SelfArx}
\usepackage[english]{babel} 
\usepackage{amsmath}
\usepackage{amssymb}
\usepackage{graphicx}
\usepackage[dvips]{epsfig}

\definecolor{color1}{RGB}{0,0,90} 
\definecolor{color2}{RGB}{0,20,20} 

\newcommand{\beginsupplement}{%
\setcounter{table}{0}
\renewcommand{\thetable}{S\arabic{table}}%
\setcounter{figure}{0}
\renewcommand{\thefigure}{S\arabic{figure}}%
}

\PaperTitle{Task-related edge density (TED) -
a new method for revealing large-scale network formation in fMRI data of the human brain}

\Authors{Gabriele Lohmann\textsuperscript{1,2}*, 
Johannes Stelzer\textsuperscript{1,2},
Verena Zuber\textsuperscript{3},
Tilo Buschmann\textsuperscript{4},
Daniel Margulies\textsuperscript{5},
Andreas Bartels\textsuperscript{6},
Klaus Scheffler\textsuperscript{1,2}}
\affiliation{\textsuperscript{1}\textit{University Hospital, Department of Biomedical Magnetic Resonance Imaging, T{\"u}bingen, Germany}}
\affiliation{\textsuperscript{2}\textit{Magnetic Resonance Centre, Max-Planck-Institute for Biological Cybernetics, T{\"u}bingen, Germany}}
\affiliation{\textsuperscript{3}\textit{Institute of Clinical Medicine, University of Oslo, Norway}}
\affiliation{\textsuperscript{4}\textit{Fraunhofer Institute for Cell Therapy and Immunology, Leipzig, Germany}}
\affiliation{\textsuperscript{5}\textit{Max-Planck-Institute for Human Cognitive and Brain Sciences, Leipzig, Germany}}
\affiliation{\textsuperscript{6}\textit{CIN Vision \& Cognition Group, Centre for Integrative Neuroscience, T{\"u}bingen, Germany}}
\affiliation{*\textbf{Corresponding author}: lohmann@tuebingen.mpg.de}

\Keywords{hallo}

\Abstract{The formation of transient networks in response to external stimuli or as a reflection of internal
cognitive processes is a hallmark of human brain function. However, its identification in fMRI data
of the human brain is notoriously difficult.
Here we propose a new method of fMRI data analysis that tackles this problem by
considering large-scale, task-related synchronisation networks.
Networks consist of nodes and edges connecting them, where nodes correspond to voxels in fMRI data,
and the weight of an edge is determined via task-related changes in dynamic synchronisation
between their respective times series.
Based on these definitions, we developed a new data analysis algorithm that
identifies edges in a brain network that differentially respond in unison to a task onset
and that occur in dense packs with similar characteristics. Hence,
we call this approach ``Task-related Edge Density'' (TED).
TED proved to be a very strong marker for dynamic network formation that easily lends itself to statistical
analysis using large scale statistical inference.
A major advantage of TED compared to other methods is that it
does not depend on any specific hemodynamic response model, and it also
does not require a presegmentation of the data for dimensionality reduction as it can handle
large networks consisting of tens of thousands of voxels.
We applied TED to fMRI data of a fingertapping task provided by the Human Connectome Project.
TED revealed network-based involvement of a large number of brain areas that evaded detection
using traditional GLM-based analysis. We show that our proposed method provides an entirely new window into
the immense complexity of human brain function.}

\begin{document}
\maketitle
\flushbottom
\thispagestyle{empty}

\section{Introduction}

The human brain is a large-scale network consisting of approximately 85~billion  neurons
that form a vast number of subnetworks on all spatial scales \cite{Azevedo:2009eg}. The properties of the intrinsic connectivity,
such as small-worldness \cite{Watts:1998db}, make possible the coexistence between local processing of information
in specialised circuits and large-scale integrative processes, involving multiple remote sites.
It has been suggested that the neuroanatomical architecture itself gives rise to a rich dynamic repertoire
setting the frame for a large number of flexibly accessible brain functions \cite{Deco:2012jv,Deco:2012ib,Honey2007}.

Traditionally, brain mapping techniques using functional magnetic resonance imaging (fMRI) have focused on studying
brain areas separately in a voxel-by-voxel fashion (i.e. univariately). The key idea behind such approaches was
to identify task- or stimulus-related changes of the blood-oxygen-level dependent (BOLD) signal activity on the local level.
The most prominent example is statistical parametric mapping using the general linear model \cite{Buxton2004,Friston1995}.
Such mass-univariate approaches  treat voxels \emph{independently} from each other \cite{Lohmann:2013kc},
however, they do not capture processes relating to integration and functional interplay
between remote brain regions~\cite{Horwitz2014,Sporns2014,Pessoa2014}.

As a result, the focus of the neuroimaging community has  shifted away from the pure segregationist
to a more integrative (i.e.  network-based)
view~\cite{Worsley1998,Carbonell2009,Friston2011,Smith2012,Hutchison2013,Fornito2013,Lohmann2013,Sporns2014}.
In the following, we give a brief overview over existing methods that go beyond traditional GLM-based activation maps.

\begin{figure*}
 \begin{center}
    \includegraphics[width=0.8\textwidth]{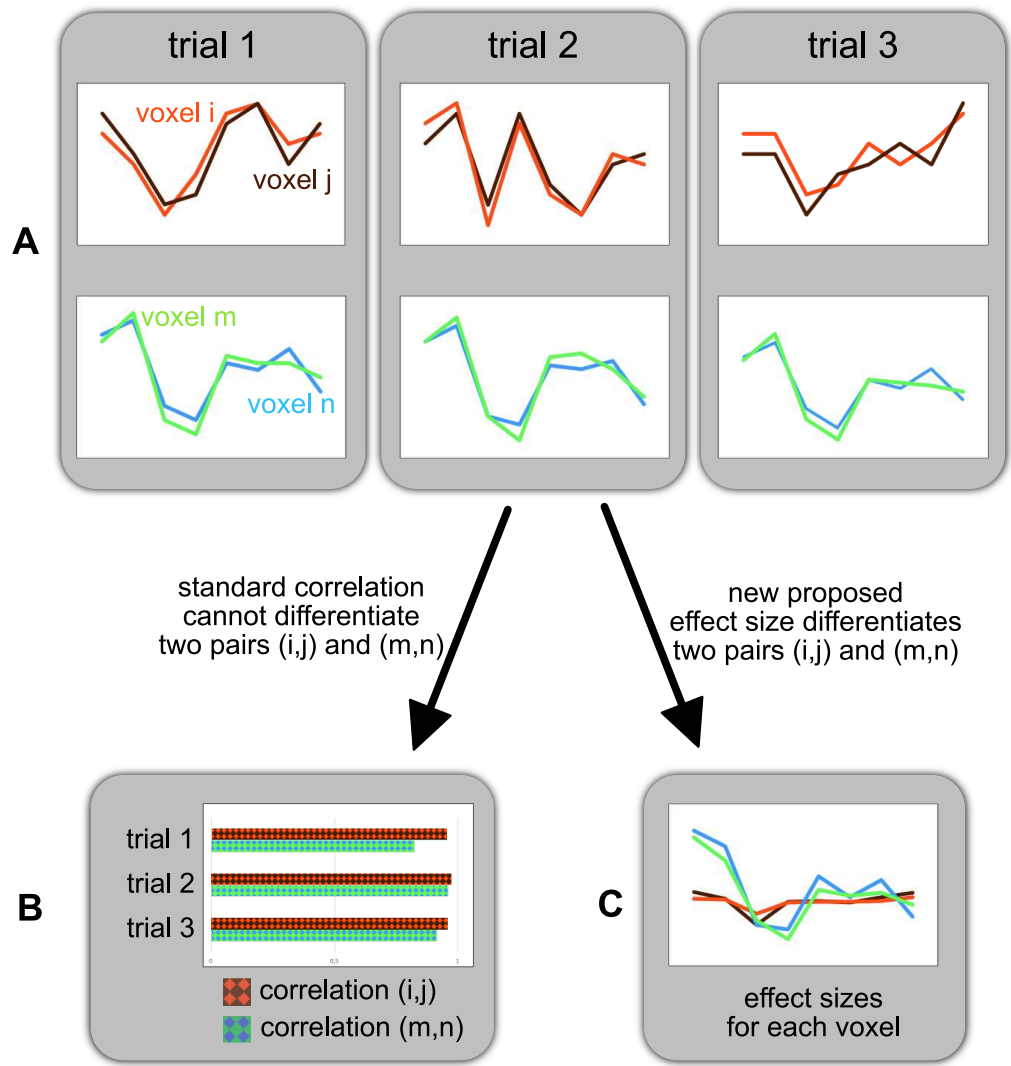}
\end{center}
\caption{{\bf Illustration of a potential problem in correlation-based statistics.}
 {(A) Hypothetical time courses of two pairs of voxels $(i,j)$ and $(m,n)$ in three experimental trials of the same condition are shown here. 
It is clearly visible that the voxel pair $(i,j)$ has a low inter-trial consistency, while the pair $(m,n)$ has a high one. 
(B) In standard correlation-based statistics, the correlation between pairs of voxels is computed for each trial. 
Thus all voxel pairs receive high values of correlation and may thus be interpreted as belonging to the same network. 
Inter-trial consistency is not taken into account here. 
However, if the connectivity between voxels is truly induced by the task and the response processing is assumed to be similar, 
the high correlations of the pair $(i,j)$ might be caused by nuisance, while the pair $(m,n)$ reflects a true effect. 
Clearly, this is not visible using a standard correlation approach. 
(C) We propose a new measure of synchronization based on effect sizes, taking into account the inter-trial consistency. 
Our measure is able to separate between the voxel pairs $(i,j)$ and $(m,n)$; the voxel pair with low inter-trial consistency receives low scores.
  }}
  \label{corrstat}
\end{figure*}

Seed-based approaches investigate how the statistical dependency between a seed voxel or area changes with respect to the rest of the brain.
The most prominent examples are correlation-based approaches~\cite{Biswal:1995tw}, where the correlation between
the time series of the seed area to all other voxels is computed. A widely used method is the \emph{psycho-physiological interaction} (PPI)
method~\cite{Friston:1997dx} and its generalisation~\cite{McLaren:2012il}, where the interaction usually is computed after
deconvolution of the fMRI signal into the neural space~\cite{Gitelman:2003hq}.
Another prominent example is the \emph{beta correlation method}
which detects the correlation of parameter estimates from the seed area to the rest of the brain~\cite{Rissman:2004hz}.
The parameter estimates themselves are derived from a general linear model (GLM).

The weak point of seed-based methods is their inability to reveal \emph{global} changes of functional reorganisation.
Only  differences \emph{relative to the seed area} can be depicted so that just a small part of the picture is revealed.
Thus, a full exploration would require a multitude of seed-based analyses (i.e. one for each grey matter location) and
to combine the resulting maps in a second step. It is easy to see that such a procedure constitutes a daunting multiple comparisons problem,
which ultimately renders a whole-brain approach infeasible. A further problem arises from analysing \emph{correlations} in time series,
as differences in correlations are in general not very reliable indicators of membership in a network.
Indeed, some times series may show very similar correlations and yet belong to very different networks. For a graphical
illustration of the issue, see Fig.~\ref{corrstat}.

The choice of seed areas raises further issues, as only few seed locations can be studied without a proper multiple comparisons correction.
Researchers thus are required to carefully select the locations in question. It is common practice to use seed locations of
special anatomical interest, or alternatively, to choose seed locations of activation peaks (e.g.~as determined by a prior
whole-brain GLM analysis).
However, the latter procedure assumes that brain regions featuring relevant changes in terms of connectivity are indeed also activation peaks.
This assumption has recently been challenged empirically~\cite{Gerchen:2014em}.

An alternative way of performing network-based analyses is to use parcellation schemes, reducing the number of network nodes.
For instance, \emph{condition-specific networks}~\cite{Dodel:2005ba} reveals changes in the whole-brain connectivity structure
that occur as response to a task, depicting the variation of functional connectivity between pairs of regions.
Similarly, \emph{network-based statistics}~\cite{Zalesky:2010iy} evaluates changes in the network structure and
incorporates a solution to the multiple comparisons problem based on a graph-based connected components methodology.
Thus, this method principally allows a larger number of smaller regions. Further methods include the adaptation of
PPI on parcellations~\cite{Gerchen:2014em}, which enables the investigation of global changes.
At the other end of the spectrum in terms of involved brain regions stands \emph{dynamic causal modelling (DCM)},
which attempts to investigate causal influences within very small networks~\cite{Friston03}. DCM's validity was challenged in~\cite{Lohmann:2012im}.

Parcellating the brain into regions comes with a number of issues, however. The choice of the parcellation scheme
underlies a certain degree of arbitrariness and is rarely motivated by anatomical features such as cyto- or myeloarchitecture~\cite{Turner2014}.
This gives rise to nonlinear properties, where small deviations in the size of regions can result in large changes in
underlying network connectivity~\cite{Liu:2011gy}. Therefore it is no surprise that the choice of parcellation scheme and thus
the number of regions have a substantial impact on the resulting network metrics~\cite{Wang:2009iz,Hayasaka:2010ey,Zalesky:2010fy,deReus:2013du}.
Furthermore, region-based approaches assume functional homogeneity within the regions~\cite{Stanley:2013ga}.
This is particularly troubling if the regions are large enough so that they can be further subdivided into parts that
feature heterogeneous connectivity profiles (e.g.~see~\cite{SolanoCastiella:2011id,Cavanna:2006eh,Thuret:2014ch,Cieslik:2013dc,Zhang:2014im}).
Averaging within such heterogeneous regions may effectively hamper the detection of subtle connectivity changes
that occur only in a subregion.
Furthermore, it should be mentioned that adopting a parcellation scheme also implies
that it is not possible to quantify the total number of connections between regions~\cite{Moussa:2011du}. Therefore,
it has been suggested that voxel-level approaches in the context of network analysis are preferable~\cite{Stanley:2013ga}.

Several other algorithms target only network hubs rather than entire networks, e.g.~\cite{Buckner2009,Lohmann10a}
and are therefore not comparable to the present approach.
Other methods such as multivariate pattern analysis  (MVPA)~\cite{Norman2006} or
independent component analysis (ICA)~\cite{Beckmann05a} also fall into a different domain, and are therefore
not discussed here.

The publications listed above all contributed immensely towards a network-based understanding of brain function.
However, they all suffer from some limitations, e.g.~they require a presegmentation of the data,
or they do not offer a mechanism for statistical inference, or they depend on a particular hemodynamic model.
The dependence on a hemodynamic model was found to be problematic in a recent study by
Gonzalez-Castillo et al. who tested a range of different hemodynamic response models and found
wide-spread activations which had previously evaded detection~\cite{GonzalezCastillo:2012js}.
They ascribed the sparsity of classical activation maps to high noise levels and overly strict response models.

Therefore, our goal in this paper is to establish a new method for fMRI data analysis
that fulfills the following requirements. It should
\begin{enumerate}
\item
identify task-related changes in network configuration,
\item
not require any presegmentations,
\item
be free from any specific hemodynamic response model,
\item
and incorporate rigorous statistical inference.
\end{enumerate}

To achieve this goal, we characterise functional networks as large-scale, task-related \emph{collective synchronisations} of the
BOLD signal measured at voxel-level resolution.
At the heart of our method is the concept of spatially localised and task-related edge density
motivating us to call this algorithm "TED" (Task-related Edge Density).
In short, TED identifies edges in a brain network that differentially respond in unison to a task onset
and that occur in dense packs with similar characteristics.
We found TED to be a very strong marker for dynamic network formation that easily lends itself to statistical analysis
using large scale statistical approaches.

In the following, we describe our proposed algorithm and demonstrate its applicability for dynamic network discovery
in task-based fMRI data provided by the Human Connectome Project (HCP)~\cite{HCP}.

\section{Materials and Methods}

Networks consist of nodes which are interconnected by edges. We define nodes as voxels and the weight of an edge between
any pair of voxels as \emph{task-related changes in dynamic synchronisation} between
their respective times series. 

{Our algorithm supports experiments presented in a block design with non-overlapping trials
of sufficient length to allow for a connectivity analysis. 
It computes a change in synchronization between two experimental conditions and it requires several repetitions 
of trials of both conditions to permit a valid statistical inference.}

The algorithm proceeds in six steps. First, the data are preprocessed using a
standard preprocessing pipeline which must include a correction for baseline drifts.
Second, we define a measure $z$ of task-related differential synchronisation for each edge in the network.
In a third step, the $z$-values are normalised.
Fourth, a measure called ``edge density'' ($D_e$) is computed for each edge,  after which edge densities
are subjected to a statistical inference procedure to assess which edges are significantly affected
by the experimental task. Finally, we propose methods for visualising the results.
In the following, we will describe the six processing steps of TED in more detail.

\subsubsection*{Step 1. Preprocessing.}
The TED algorithm assumes that the fMRI data have been preprocessed using some standard preprocessing
pipeline. This should generally include corrections for motion, slicetiming, and EPI-related distortions
as well as a removal of baseline drifts. In the case of multi-subject studies, a geometric alignment with
the MNI anatomical template is needed.
{Physiological noise removal should be included into the preprocessing chain if there is reason
to assume that it differentially affects the two task conditions.}

\subsubsection*{Step 2. Obtaining a measure $z$ of task-related differential synchronisation.}

\begin{figure*}
 \begin{center}
  \includegraphics[width=1\textwidth]{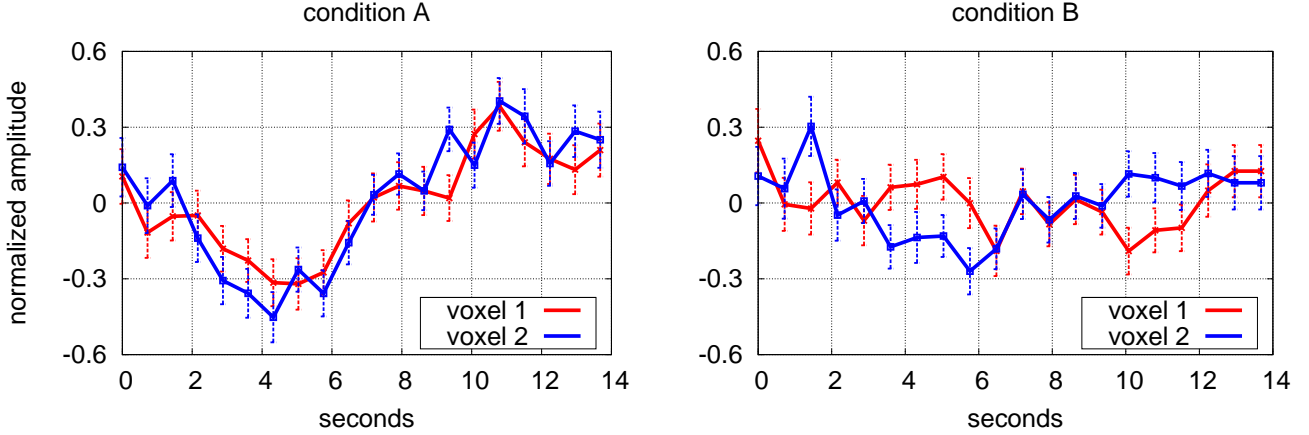}
 \end{center}
  \caption{{\bf Illustration of differential synchronisation $\theta_{i,j}$}.
    The figure shows mean $\mu(t)$ and standard errors $\sigma(t)$ across 100~trials in a pair of voxels $i$ and $j$.
    The left pane shows experimental condition $A$, the right pane condition $B$.
    The two voxels of condition $A$ appear to be stronger synchronised than those of condition $B$.
    This is reflected by a high value of synchronisation in condition A which is $\theta^A_{i,j}=1.453$
    while for B it is only $\theta^B_{i,j}=0.108$.
    The time courses are taken from two voxels of the experimental data described in this article.
  }
  \label{diffsync}
\end{figure*}

Let $A$ and $B$ denote two experimental conditions such as left hand versus right hand fingertapping
presented in a block design.
{In our experiments, the trial duration was 12~seconds.}

For simplicity, we assume that all trials have the same duration $T$, and there are $K$ number of trials per condition.
For condition $A$, let $v_i^A(k,t)$ denote the time course of voxel $i$ of trial $k$ at time $t$.
We now define a measure that quantifies the amount of
task-related change in connectivity between any two voxels $i$ and $j$.
A seemingly straightforward way, such as correlation-based statistic (CBS),
would be to simply compare the linear correlations
of their respective time series during the execution of different tasks.
However, CBS can be problematic because it assumes that similarities in correlations are
sufficient for assuming membership in common network, which is problematic (see Fig.~\ref{corrstat}).
In the following, we therefore propose a different measure which we call {\em differential synchronisation} $z$.

We first compute the average $\mu$ and standard deviations $\sigma$ across all trials as follows.

\begin{equation}
\label{def1}
\mu_i^A(t) = \frac{1}{K}\sum_{k=1}^K v_i^A(k,t)
\end{equation}
\begin{equation}
\label{def2}
\sigma_i^A(t) = \sqrt{\frac{1}{{K-1}} \sum_{k=1}^K \left(  v_i^A(k,t) - \mu_i^A(t) \right) ^2}
\end{equation}

For each voxel $i$ we thus obtain an effect size at time point $t$:
\[
s_i^A(t) = \frac{\mu_i^A(t)}{\sigma_i^A(t)}
\]

The synchronisation $\theta^A_{i,j}$ between voxels $i$ and $j$ in condition $A$ is then defined as
the z-transformed linear correlation between $s_i(t)$ and $s_j(t)$. More precisely, we have

\[\theta^A_{i,j}  = \left\{
\begin{array}{ll}
\frac{1}{2} \log \frac{1+r^A_{i,j}}{1-r^A_{i,j}} &  $for$ \; r^A_{i,j} > 0  \\
0 &  $otherwise$\\
\end{array}
\right.
\]
{with $r^A_{i,j}$ denoting the Pearson correlation coefficient between $s^A_i$ and $s^A_j$.}

The synchronisations for experimental condition $B$ are computed analogously.

We now have two $n \times n$ matrices $\Theta_A$ and $\Theta_B$
each recording the task-related synchronisation in conditions $A,B$ for all pairs of voxels $i,j=1,...n$ where $n$
is the number of voxels. Based on these two matrices, we
introduce a measure of differential synchronisation $z$ defined as an elementwise subtraction:
\begin{equation}
\label{eq1}
z_{i,j} = \theta_{i,j}^A - \theta_{i,j}^B
\end{equation}
Note that large positive values of $z$ indicate a higher synchronisation in condition $A$ compared to condition $B$,
see Fig.~\ref{diffsync} for an illustration.

{Negative correlations are excluded in the definition of $\theta$ to avoid misinterpretations.
Specifically, consider a case where two voxels are not correlated at all in experimental condition $A$,
while showing a strong negative correlation in condition $B$. If the synchronisation $\theta$ were allowed to take negative values, 
then this would entail $z = \theta^A - \theta^B \gg 0$ which might be mistaken for a task-positive involvement
of the connection between these two voxels in condition $A$, even though the correlation is in fact absent.}

Also note that $z$ is designed to enforce task-related synchrony across trials within the same experimental condition
so that the problem illustrated in Fig.~\ref{corrstat} does not arise.

\subsubsection*{Step 3: Normalisation of $z$-values.}
The output of the previous step is an $n \times n$ symmetric matrix of $z$-values.
With $n \approx 54,000$, this matrix has more than one billion entries.
Theoretically, we might want to apply a statistical test to the elements of this matrix
in order to find significant task-related differences in synchronisation between two voxels.
However, such an approach would be problematic because the  $z$-values must be expected to
be heavily influenced by non-neuronal confounds such as
cardiac and respiratory effects or subject motion~\cite{Power2012,Khalili-Mahani2013}.
Furthermore, it is well known that neuromodulatory effects (state of arousal, attentiveness, etc.)
have a major impact on the fMRI signal~\cite{Logothetis2008}.
It is extremely difficult to disentangle ``true'' neuronal effects from the confounding effects listed above.
Therefore - instead of trying to solve this challenging problem - we propose to evade it using
normalisation. Specifically, we choose the following approach.

We begin by computing the probability density distribution of the $z$-values.
Theoretically - if no effect were present in the data - this distribution should be a
Gaussian normal. However, due to confounds, we cannot expect this to be the case.
Therefore, we apply a histogram matching procedure so that this distribution does indeed
follow a Gaussian normal with mean zero and standard deviation one. We call this procedure
{\em $z$-normalization} and it can be easily achieved by any standard histogram matching
algorithm~\cite{Gonzalez2006}.

At this point, it would not make sense to apply an elementwise statistical test to the normalized
$z$-values to check for significant differences from zero. The reason is that after normalization
the distribution of the $z$-values is exactly the same as that of the theoretical null distribution.
In other words, normalization effectively eradicates elementwise effects that might have been present in the
$z$-values - regardless of whether or not they were neuronal or non-neuronal in origin.
Clearly, this step makes our approach very conservative. But since normalization is a monotone transformation
it preserves the ranks of the $z$-values, and it also preserves  spatial information so that
small neighbourhoods containing mostly high ranking $z$-values will also have high ranking
normalized $z$-values. In the following, we will solely exploit information of this type.

\subsubsection*{Step 4: Edge densities.}

\begin{figure*}
\begin{center}
  \includegraphics[width=0.8\textwidth]{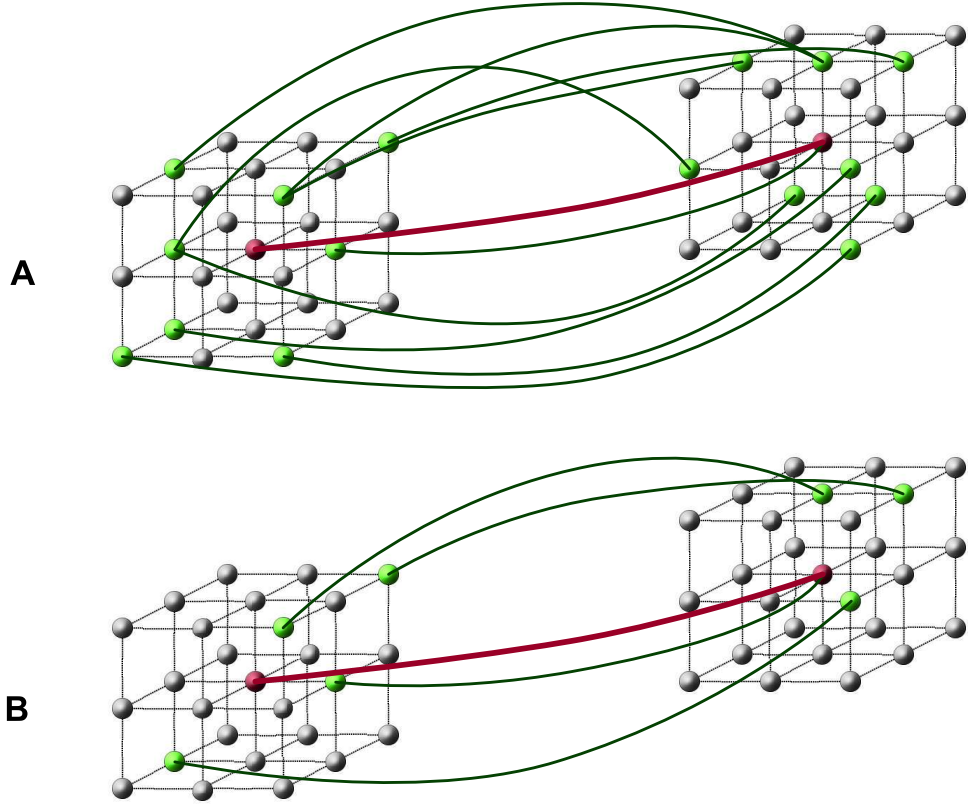}
\end{center}
  \caption{ {\bf Schematic illustration of edge density ($D_e$), showing a case with high $D_e$ and one with low $D_e$}.
   { In this figure, voxels are depicted as small spheres. The value of $D_e$ of the (thick red) edge connecting the
    two blue voxels in the centre is computed as follows:
    First, we consider the 26-adjacent neighbours of its endpoints (shown here as grey and green spheres). 
    Theoretically, the highest possible number of edges connecting any two endpoint voxels \emph{across} the 26-neighbourhoods
    is $27 \times 27 = 729$. We define the edge density $D_e$ as the number of edges whose $z$-values are above
    a user-defined threshold $z_t$ divided by the total number of possible edges (i.e. 729). 
    In the above examples, lines connecting nodes (voxels) indicate supra-threshold edges. 
   In example (A), 11 out of 729 possible edges are above threshold, thus $D_e = \frac{11}{729} \approx 0.015$. 
   In example (B), only 5 out of 729 possible edges are above threshold, thus, $D_e = \frac{5}{729} \approx 0.007$.}}
  \label{ED}
\end{figure*}

{
We now propose a new network metric that draws on spatial adjacency as the key source of
information. We call this feature ``edge density'' ($D_e$). The edge density is computed for all edges in the graph that surpass an initial 
user-defined threshold $z_t$. 
The value of $D_e(i,j)$ for an edge connecting two spatially separate voxels $i,j$ indicates to what degree the two neighborhoods of 
the voxels $i$ and $j$ are connected with each other. A high edge density indicates that many edges connect the two neighbourhoods, 
while a low edge density indicates that only few links are present, for an illustration see Fig.~\ref{ED}.
}
 
{
Quantitatively, the edge density is defined as follows: first, the total number of possible edges between the neighborhoods of 
the voxels $i$ and $j$ is computed (omitting local connections, i.e. the start and ending point of an edge must be in different neighborhoods). 
Next, the number of supra-threshold edges between the neighborhoods is determined, only taking into account edges whose 
normalised $z$-value exceed the threshold $z_t$. The fraction between this number of suprathreshold edges and 
the total possible number of edges then defines the edge density $D_e$. Thus, if all neighbouring edges have 
supra-threshold $z$ values, $D_e$ will be one, and if most edges fall below the threshold, $D_e$ scores will approach zero. 
To summarize, $D_e$ indicates to what degree the local neighbourhoods of pairs of voxels show a similar change of connectivity 
across the experimental conditions.
As the edge density is only computed for edges whose $z$ is larger than $z_t$, all other edges are defined to have an $D_e$ of zero.
In our experiments, we considered the top 1~percent of all edges so that $z_t$ was set to 2.33.
Note that the computation of $D_e$ requires a specification of adjacency.
In our experiments, we used 26-neighbourhoods, but 18- or 6-adjacencies may also be considered.
}

{
Also note that we exclude short edges from further analysis.
A short edge is an edge whose endpoints $i,j$ have a Euclidean distance of less that 15mm.
The reason for doing this is that the two neighbourhoods should be non-overlapping, and because of spatial
smoothness we additionally increased the minimum distance so that the borders of the two neighbourhoods are at least
three voxels apart.
}

\subsubsection*{Step 5: Statistical inference.}
In order to correct for multiple testing we employ a procedure controlling the false discovery
rate (Fdr)~\cite{Benjamini95,Efron2007,Strimmer08}.

{
The original Fdr algorithm proposed in~\cite{Benjamini95} requires that data points are independent
and that the null distribution is uniform. Both requirements may not be not fulfilled in our case. 
Therefore, we use a different formulation of Fdr that is well suited for large-scale
statistics involving a large number of data points with complex dependencies among them~\cite{Efron2007,Strimmer08}}.

The basis of {this} Fdr approach is the assumption of a two component mixture model
for the ${D_e}$-scores based on a null and non-null component with a cumulative distribution function (cdf)
$F_0({D_e})$ and $F_1({D_e})$ respectively. Additionally, we define an a priori probability for being null
by $\pi_0$. Then, the observed joint density $F_{joint}({D_e})$ is given by

\begin{equation}
F_{joint}({D_e}) = \pi_0 \times F_0({D_e}) + (1-\pi_0) F_1({D_e})
\end{equation}

The Fdr is defined as the probability of being null, or a false discovery, given a ${D_e}$-score as big or bigger
than the observed one. This translates into

\begin{equation}
Fdr({D_e}) = \pi_0 (1-F_0({D_e})) / (1-F_{z}({D_e})).
\end{equation}

For simplicity we assume $\pi_0=1$, which is the most conservative choice.
The cdf $F_{joint}$ is estimated from histogram counts of the edge densities using cumulative summation.
As we do not have evidence for a theoretical null distribution we rely on an empirical permutation null estimate,
where the null cdf $F_0$ is estimated by using random permutation of task labels~\cite{Nichols:2002vw}.
More precisely, we randomly construct a binary permutation vector $\rho$ of size $K$ where each entry $k \in {1,...K}$ indicates
whether or not the task label should be swapped in trial $k$. We used Bernoulli random trials with probability $p=0.5$ for this purpose.
This yields

\[\rho(k) = \left\{
\begin{array}{lr}
0 & : $ k-th trial original  $\\
1 & : $ k-th trial swapped$
\end{array}
\right.
\]

The permuted group mean amplitudes for the experimental conditions $A$ and $B$ for voxel $i$ then are defined as

\[
\mu_i^{'A}(t) = \frac{1}{K}\sum_{k=1}^K {(1-\rho(k))} v_i^A(k,t) + \rho(k) v_i^B(k,t)
\]

\[
\mu_i^{'B}(t) = \frac{1}{K}\sum_{k=1}^K \rho(k) v_i^A(k,t) + {(1-\rho(k))} v_i^B(k,t)
\]

with the standard deviations for experimental condition $A$ and $B$
\[
\sigma_i^{'A}(t) = \sqrt{\frac{1}{K} \sum_{k=1}^K \left[  {(1-\rho(k))} v_i^A(k,t) + \rho(k) v_i^B(k,t)  - \mu_i^{'A}(t) \right] ^2}
\]

\[
\sigma_i^{'B}(t) =  \sqrt{\frac{1}{K}  \sum_{k=1}^K  \left[ \rho(k) v_i^A(k,t) + {(1-\rho(k))} v_i^B(k,t) - \mu_i^{'B}(t) \right] ^2}
\]

In other words, for the special case of the permutation vector $\rho(k) = 0$, $k=1...K$, the above definitions simplify
to the original definitions (\ref{def1}) and (\ref{def2}).

In order to ensure that spatial smoothness is preserved we apply the same permutation vector $\rho$
to all voxels in the brain mask. Therefore, differences between $F_0$ and $F_z$
cannot be attributed to the spatial correlations that are generally inherent in fMRI data.
As described in more detail below, a relatively small number of permutations may suffice to converge to a stable
estimation of $F_0$, and hence of the false discovery rate.

\subsubsection*{Step 6: Visualisation of results.}
The previous processing step yields a set of edges that indicate significant changes in task-related connectivity.
Since the number of such edges can be very large, visualisation of results may become difficult.
Here we propose two different methods.

The first method is to project edges onto a ``hubness map''. A voxel in the hubness map
records the number of edges for which this voxel serves as an endpoint.
Voxels in which many edges accumulate may be viewed as hubs in a task-specific network,
and the number of edges meeting in a voxel is a measure of the voxel's hubness, see Fig.~\ref{hubs_motor}.
Note however, that this measure of hubness should not be confused with activation strength
as can be seen from figure~\ref{diffsync}. Here differential synchronisation goes along with a decrease in BOLD activation
rather than an increase.

The second method is to display edges as lines in a 3D~rendering such that each line represents an edge
that survived significance thresholding. Such renderings can become quite cluttered and therefore
edges that are close to each other are bundled together to produce a clearer picture~\cite{Boettger2013}.
We use the software package ``braingl'' for this purpose~\cite{braingl_code}, see Fig.~\ref{tedroi}.

\subsection*{Experimental Data}

We applied TED to task-based fMRI data provided by the Human Connectome Project (HCP),  WU-Minn Consortium~\cite{HCP,Barch2013}.
We focused on the motor/fingertapping task using minimally preprocessed fMRI data of 100~participants.
The preprocessing protocol is described in~\cite{HCPpreproc}.
The experiment was acquired in two separate runs with one run using left-right phase-encoding,
and the other run using right-left phase-encoding.
While in the scanner, participants were cued visually to tap
their left or right fingers, squeeze their left or right toes, or move their tongue. Each block lasted
12~seconds (10 movements),
and was preceded by a 3~second cue.
{Since the repetition time was 720 milliseconds,
there were 16.666 volumes (time steps) per trial of which we used the initial 16.}

In each of the two runs, there were 13~blocks, with 2 of tongue movements, 4 of hand movements
(2~right and 2~left), 4~foot movements and three 15~s fixation blocks per run, see~\cite{Barch2013}.
Here, we only used the second of the two fingertapping blocks
so that we have 100~trials for each condition $A$ and $B$ (right hand tapping and left hand tapping)
in each phase-encoding run.

In order to reduce the number of voxels and hence the computational load, we downsampled the data
to isotropic voxels of size $(3.0mm)^3$ from the original resolution of $(2.0mm)^3$.
We also corrected for baseline drifts using a highpass filter
with a cutoff frequency of 1/90 Hz. To reduce the effect of anatomical variability across subjects,
we applied spatial smoothing with a Gaussian smoothing kernel of FWHM=5mm.
Spatial smoothing is however not an integral part of the algorithm
and should be omitted whenever possible, particularly for single-subject analysis~\cite{Stelzer2014}.
We manually defined a region of interest (ROI) containing
about 54,000 voxels covering the entire brain including grey and white matter, subcortical
structures, CSF and the cerebellum.
Furthermore, we normalised the voxel-wise time series of each trial, so that
their time series have mean zero and standard deviation one.

We performed the initial three steps of TED separately for each of the two phase-encoding runs
resulting in two matrices of normalised $z$-values. These two matrices were then combined
via a conjunction analysis by taking the element-wise minimum of the $z$-values.
For the subsequent edge density computation we used a threshold of $z_t = 2.33$.

We applied TED as described above using 1000~random permutations to estimate the null distribution.
The TED algorithm is implemented in C/C++ and makes use of parallel computation.
With 54,000 voxels TED requires about 13~GByte of main memory, and
one permutation takes about three minutes of computation time on a Linux PC using
12~parallel cores. For the computation of all 1000~permutations we made use of the
Max-Planck-Society's high-performance computing center in Garching, Germany.

\section{Results}

\begin{figure*}
\begin{center}
\includegraphics[width=1\textwidth]{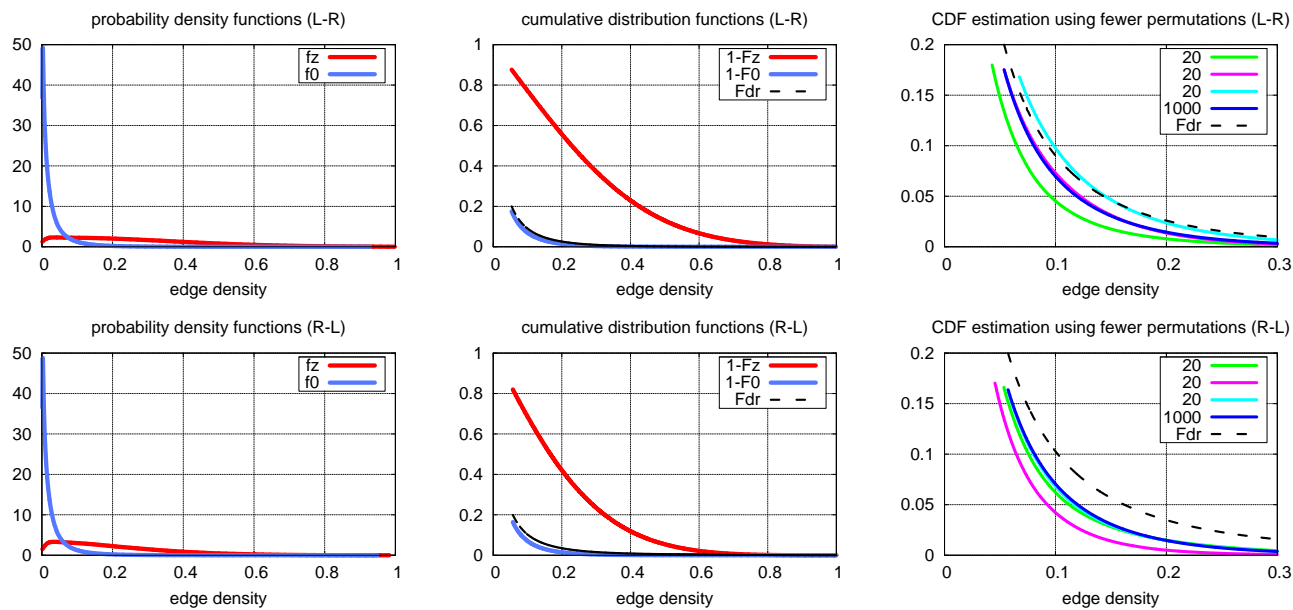}
\end{center}
\caption{ {\bf Probabilities of the edge density estimated from fMRI fingertapping data.}
The top rows contrasts left hand minus right hand tapping,
the bottom row is the reverse contrast (right hand minus left hand).
The left plots show estimations of the probability density functions $f_0$ (permutation-derived) and $f_z$ (no permutation)
using Gaussian kernels with kernel bandwidth defined via Silverman's rule~\cite{Silverman}.
The plots in the middle show the corresponding cumulative distribution functions (CDF) $F_0$ and $F_z$.
Note that the differences between the null and the ``real'' distributions are massive.
The CDFs are used to estimate the false discovery rates (Fdr) which are indicated here by dashed black lines.
For better visualization, the same Fdr curves are also shown in the plots on the right.
The Fdr is used to determine a significance threshold.  For the left minus right contrast,
the cutoff was found to be $D_e > 0.1435$, i.e. for edges with $D_e > 0.1435$ the false discovery rate falls below 0.05.
For the reverse contrast, the cutoff was  $D_e > 0.1621$.
The estimation of $F_0$ is based on 1000 random permutations.
The right plots show estimations of $F_0$ using only 20 different permutations.
Note that the plots are zoomed for better visualisation. Here we show results of
three random selections of such shorter permutations vectors.
Note that they are very similar to the estimation based on 1000 permutation leading to very
similar estimations of Fdr.
In other words, far fewer than 1000 permutations would have sufficed to reach a similar result.}
\label{fdr}
\end{figure*}

\begin{figure*}
\centerline{
\includegraphics[width=1.0\textwidth]{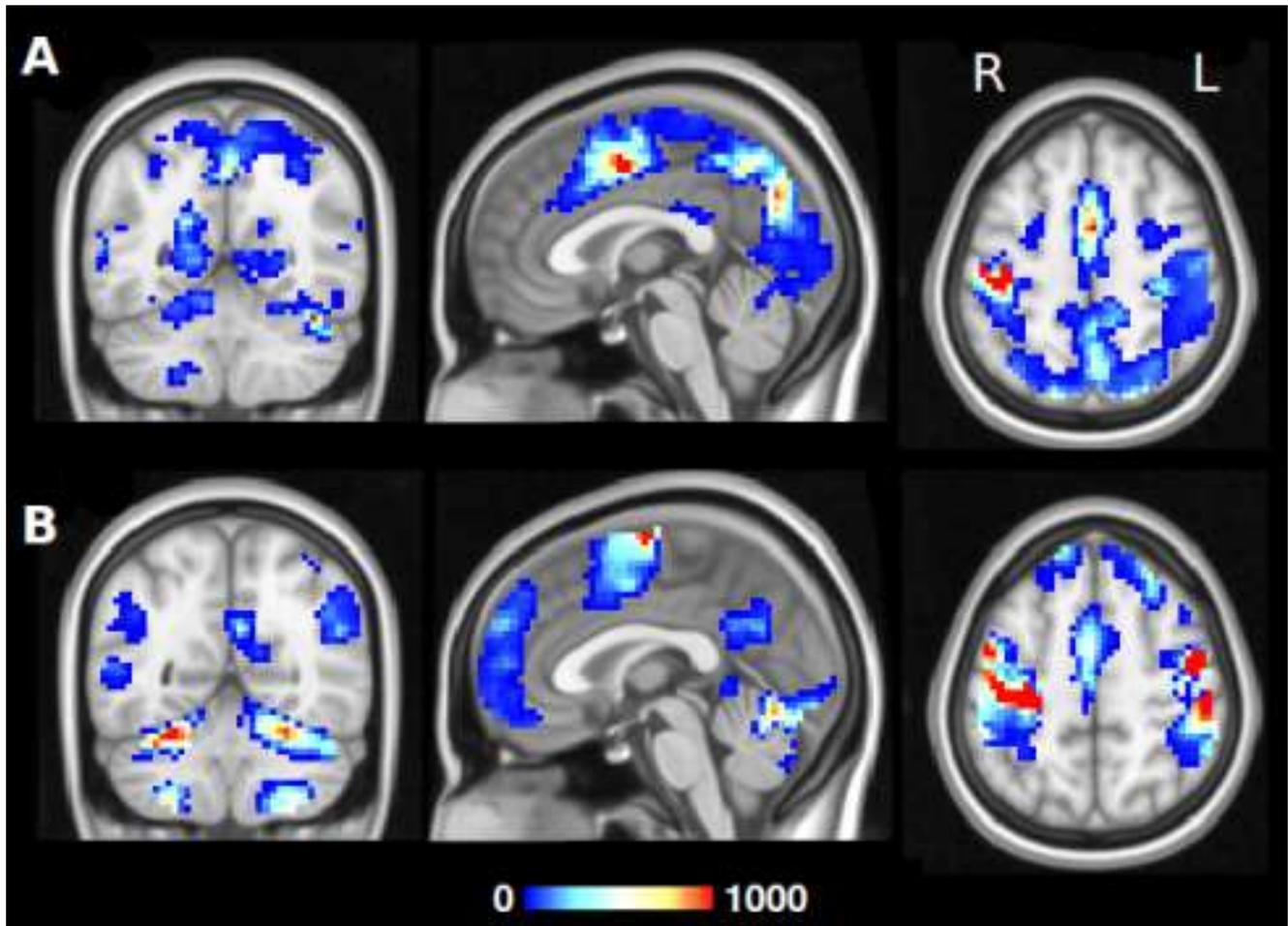}}
\caption{ {\bf Task-dependent dynamic reconfiguration of whole-brain networks}. We depict the reconfiguration using hubness maps
on basis of fMRI data fingertapping data of the Human Connectome Project. The hubness maps indicate the number of network edges that feature
a significant change between the two experimental conditions. The top row (A) contrasts
right hand minus left hand tapping, the bottom row (B) shows the reverse contrast.
The colours encode the number of edges with $Fdr < 0.05$ having one of their endpoints
in the respective colour-coded voxel and ranges from 1~to~1000. This number can be interpreted as a measure of ``hubness''.
Thus, red values in the above figure indicate hubs where many edges accumulate in a voxel.
See also supplementary Figs.~\ref{suppl_zlr},\ref{suppl_zrl}.}
\label{hubs_motor}
\end{figure*}

\begin{figure*}
\centerline{
\includegraphics[width=0.7\textwidth]{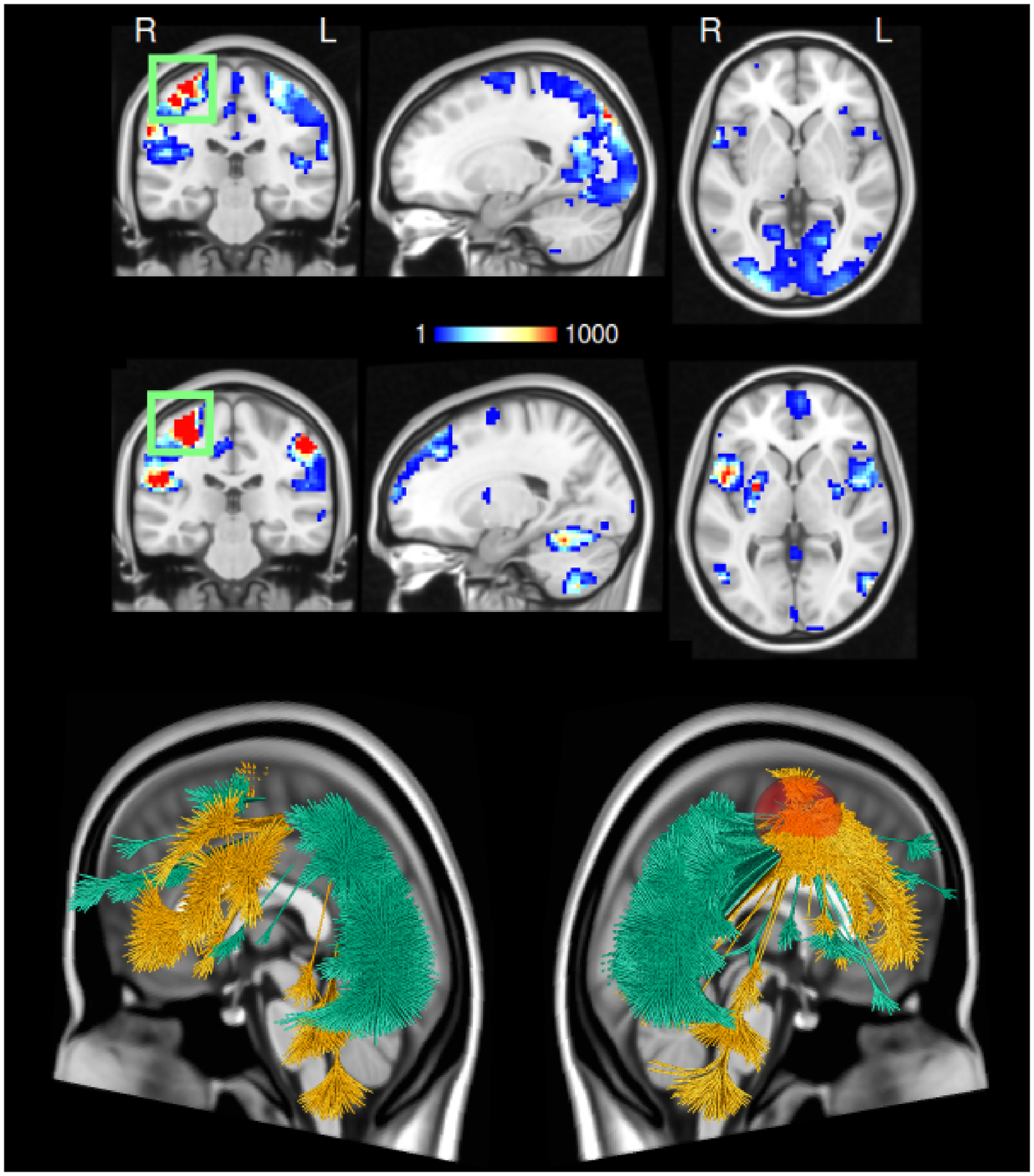}}
\caption{ {\bf Task-related reconfiguration of the right-hemispheric primary motor cortex.}
In a TED analysis, a single region may appear in opposite contrasts because
it may participate in different task-dependent networks. In this figure we show the participation of the right primary cortex
(hand knob area, marked with a green box) in different networks, depending on the experimental condition.
The exact shape of the region of interest is shown in supplementary figure~\ref{suppl_zroi}.
In the upper panel we display voxels involving all edges featuring a stronger synchronisation with the right primary motor cortex for
right hand fingertapping (as compared to left hand tapping). In the lower panel we show the voxels where the synchronisation with the
same area is higher for left hand fingertapping (as opposed to right hand tapping). The maps reveal a striking difference in synchronisation
of the right primary motor cortex to the rest
of the brain: the right hand tapping condition involves stronger synchronisation between the right motor cortex and regions in the visual
and parietal cortex. On the other hand, in the left hand condition the same area synchronizes more with bilateral areas in the cerebellum,
V5, the putamen and insular cortex and furthermore the medial prefrontal cortex.
Below, the same data are shown using a 3D rendering using the software package ``braingl''~\cite{braingl_code,Boettger2013}.
Here, the synchronisation network of the right primary motor cortex (the red sphere) in the right hand minus left hand contrast is
shown in green, the reverse contrast is shown in yellow.}
\label{tedroi}
\end{figure*}

We first contrasted global connectivity related changes for right hand minus left hand tapping.
Figure~\ref{fdr} shows the distributions $F_z$ and $F_0$ that were used to estimate
statistical significance.
At a false discovery rate of $Fdr < 0.05$ the edge density cutoff was found to be $D_e = 0.1621$, i.e.
edges with $D_e > 0.1621$ can be assumed to indicate a significantly stronger synchronisation in
left hand versus right hand tapping.
Corresponding plots for the reverse contrast can also been found in figure~\ref{fdr}.
Note that far fewer than 1000 permutations would have sufficed to reach a similar result
so that the entire computation could actually have been performed on a standard PC.

Figure~\ref{hubs_motor} shows a resulting hubness map produced as described above in step~6.
Voxels that are colour-coded are endpoints in an edge significantly affected by the task.
Voxels in which many edges accumulate
may be viewed as ``hubs'' in a task-specific network, and the number of edges meeting in
a voxel is a measure of the voxel's ``hubness''.  The upper panel indicates the hubness for edges in the contrast right hand minus left hand.
Tapping with the right hand as opposed to the left increased the global connectivity in supplementary motor areas, right and left motor cortex,
somatosensory areas, the frontal eye fields, regions in the parietal cortex and the visual cortex. On the other hand, as shown in the
lower panel, left hand tapping minus right hand tapping seemed to increase the global connectivity within the bilateral motor network,
the default mode network, bilateral putamen, bilateral V5, insular cortex and regions in the cerebellum.

\begin{figure*}
\begin{center}
\includegraphics[width=1\textwidth]{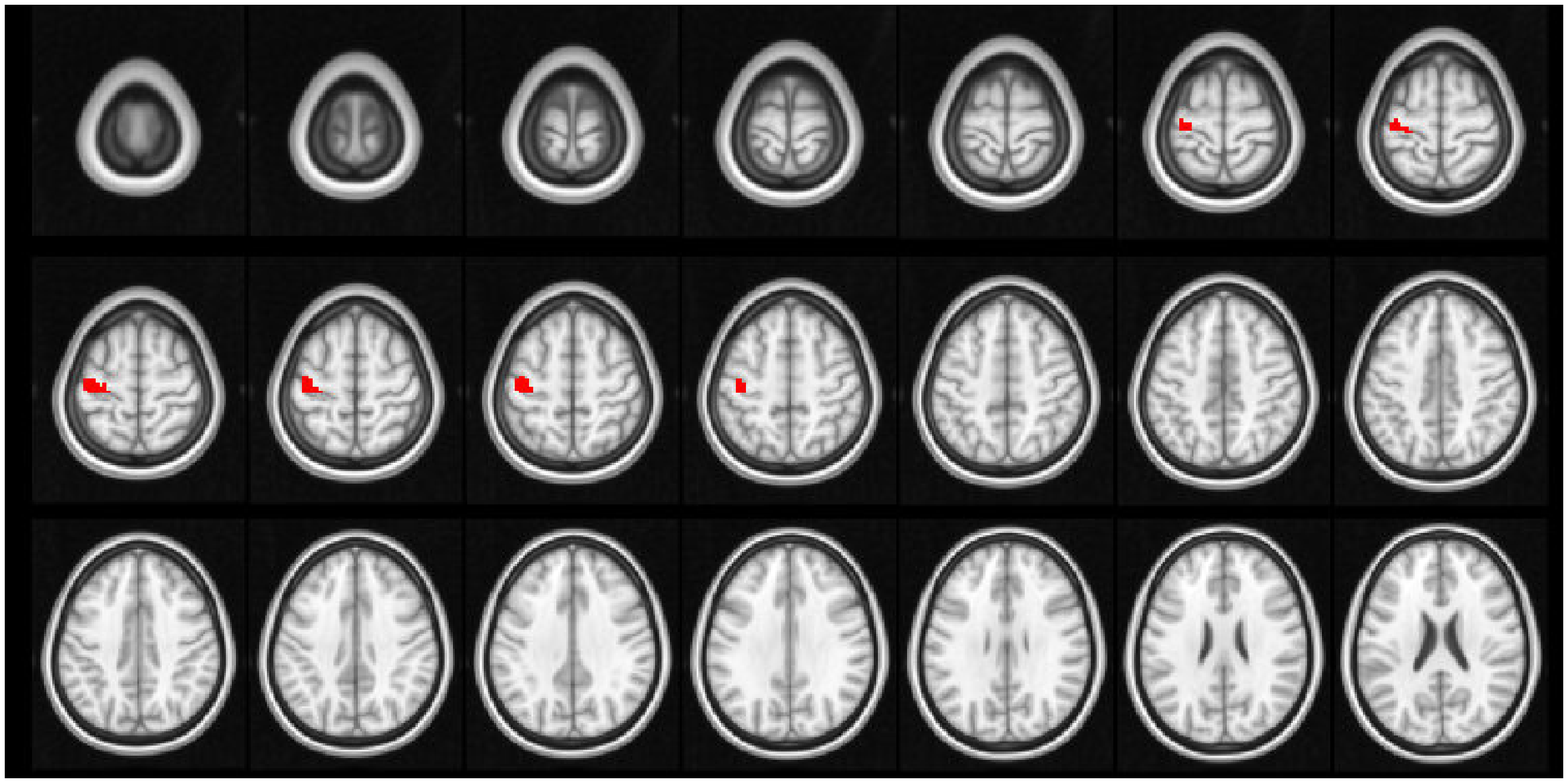}
\end{center}
\caption{{\bf Region of interest in the right motor area.}
The map shows the region of interest in the right moto area used in
figure~\ref{tedroi} of the main manuscript.}
\label{suppl_zroi}
\end{figure*}

In the same figure it can be seen that several regions appear involved in both contrasts.
The reason is that a single brain region may participate in two different
networks that are differentially activated by the tasks.
Note that such a case cannot be detected in a classical GLM-based analysis.
As an example, we investigated the right primary motor cortex which is involved in both contrasts,
see Figs.~\ref{tedroi},\ref{suppl_zroi}.
Here we only investigated edges with one endpoint in a preselected region of interest, which roughly corresponds
to the right hand area.
The maps reveal a striking difference in synchronisation of the right primary motor cortex to the rest
of the brain: the right hand tapping condition involves stronger synchronisation between the right motor cortex and regions in the visual
and parietal cortex. In the left hand condition, the same area shows higher synchrony with bilateral areas in the cerebellum,
V5, the putamen and insular cortex and furthermore the medial prefrontal cortex.

\subsection*{Comparison with a GLM-based analysis.}

\begin{figure*}
\centerline{
\includegraphics[width=0.9\textwidth]{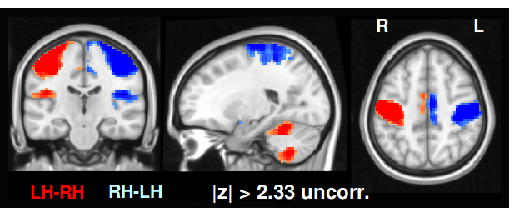}}
\caption{ {\bf Classical univariate GLM-based analysis. }For comparison to standard analysis methods, we used a univariate activation
based GLM technique (see methods). We thresholded the activation map very liberally at $|z| > 2.33$ on the voxel level
without correcting for multiple comparisons.
See also supplementary Fig.~\ref{suppl_glm}.}
\label{glm}
\end{figure*}

For comparison, we performed a standard analysis using the GLM approach as implemented in Lipsia~\cite{Lohmann:2001ww}. The preprocessing
of the data was performed as described above. We computed activation maps for each of the two phase-encoding runs separately
using the general linear model. These maps contain uncorrected $z$-values representing the contrast between left hand minus right hand
fingertapping. As in the TED approach, we performed a conjunction analysis on the two maps, where the voxel-wise minimum value
of both Z-maps was used for the case that both were positive,
the maximum value in case both were negative, and zero for diverging signs of the $z$-values. We then thresholded the resulting
conjunction map such that voxels with $|z| > 2.33$ remained. No multiple comparisons correction was applied.
The resulting map contrasting left hand minus right hand fingertapping (see Fig.~\ref{glm}) shows the voxel-wise
differences in activation strength. The activations include bilateral motor areas, pre-SMA, as well as the cerebellum.

\section{Discussion}

In this study, we introduced a new algorithm called ``TED'' which was designed to
identify task-related reconfigurations in brain networks without requiring presegmentations and
without being dependent on some specific hemodynamic response function.
Since network-based analysis methods face tremendous multiple comparison problems, statistical
inference was a key concern in this context.

At the heart of TED is the concept of ``edge density''. A voxel is deemed to be involved in the task if
it has a partner voxel at some spatially distinct location with a similar synchronicity trajectory, so that
these two voxels form an edge with a supra-threshold differential synchronisation. At the same time, such an
edge is required to appear in a dense pack of neighbouring edges with similar characteristics.
This approach freed us from the need for any explicit hemodynamic modelling, and it also allowed us
to make inferences at the spatial resolution of small neighbourhoods around individual voxels without
requiring a presegmentation.

TED's validity is tested via a comparison against a null model derived from permutation testing.
We found that the null and the real distributions of edge densities differ massively
so that statistical significance could be easily established. The amount of this difference
is especially surprising since it is based on normalized $z$ values so that
the size of this effect can only be ascribed to spatial adjacency, not to the magnitudes of the $z$
values themselves. We normalized the $z$ values to avoid potential confounds from physiological noise.
Spatial structure was preserved in our random permutations so that
our results cannot simply be due to the spatial smoothness that is generally inherent in fMRI data.

The results obtained with TED suggest a dominant role of local inter-connected neighbourhoods
forming transient task-related networks with other local neighbourhoods. In this regard,
our concept of edge density is somewhat related to the concepts of a clustering coefficient
and small-worldness~\cite{vandenHeuvel:2008gi,Watts:1998db}.
However, edge density differs from the clustering coefficient in that it measures connectivity between
two local neighbourhoods at spatially separate areas.
In the literature, there exist several approaches comparing the differences between local and global
functional connectivity profiles~\cite{Sepulcre:2010is,Tomasi:2011fv}.
It has been found that different brain regions exhibit a varying balance between such local and global connectivity,
which may further depend on the current experimental state (i.e. task). On the other hand, the results of both Sepulcre et al.~\cite{Sepulcre:2010is}
and Tomasi et al.~\cite{Tomasi:2011fv} indicate a strong overlap between regions which feature both increased local and global functional connectivity.
Our methodology offers a potentially interesting perspective on this, as according to our interpretation global changes
in connectivity may be accompanied by local interconnections, reflecting the dynamics of local subnetworks
that form transient long-range connections.

The hubness maps produced by TED are generally consistent with the standard univariate GLM analysis and also with
a meta-analysis on fingertapping~\cite{Witt2008}.
However, TED found several sites that significantly changed their task-related TED trajectories
while not reaching significant net-BOLD modulation when analysed using the traditional GLM approach.
This agrees with earlier findings by Gerchen et al.~\cite{Gerchen:2014em}.
Also, some regions with negative task involvement showed positive TED changes, and vice versa.
This shows that brain areas that appear non-significant in a GLM activation map may nevertheless
form relevant hubs in a task-positive network, and that the sign of task-related activity can be
independent of the sign of task-related (de-)synchronisation.
TED-based fMRI analysis thus constitutes a novel analysis approach that complements traditional GLM based analyses.
In particular, it can reveal task-related regional involvement that evades detection using traditional approaches.

In this respect, our work is in line with an earlier study by Gonzalez-Castillo et al. who also found large-scale
time locked activity that went undetected in the classical fMRI analysis~\cite{GonzalezCastillo:2012js}.
The authors ascribed the sparsity of traditional activation maps to high noise levels and
overly strict predictive response models.
Our present study goes beyond this earlier work in that we present data-driven criteria
that can be subjected to rigorous statistical significance testing.
Also, the amount of scan time required to achieve this result was considerably smaller than in this earlier study.
Here we only needed  $4 \times 100$ trials where each trial was 12~seconds long.
Adding a hypothetical intertrial interval of perhaps 18~seconds we arrive at a scan time of roughly 200-minutes.

In the present study we applied TED to group level data so that spatial accuracy was limited. This was even further reduced
because we had to downsample the data to $(3mm)^3$ resolution to ease the computational burden, see~\cite{Stelzer2014}
for a discussion on problems relating to spatial inaccuracies.
These limitations are not implicit in the TED algorithm, so that it is quite possible to apply
TED to single subject data at very high spatial resolutions provided sufficient computational resources exist and
a sufficient number of experimental trials are acquired to yield sufficient statistics.
Since TED specifically targets local neighbourhoods, we expect that the results will benefit from more precise
spatial information, e.g.~provided by ultrahighfield MRI~scanners.

{Note that our present implementation of TED assumes that inter-trial variance is the only source of
variation in the data.
However, the input data may be structured such that several sources of variance
need to be addressed. For instance, both within-subject and between-subject subject variance may be relevant.
A potential way of handling such cases may be to perform the first three steps of TED for each subject 
individually. This results in a separate $z$-matrix per subject each incorporating within-subject variance.
These matrices can then be subjected to a onesample t-test incorporating between-subject variance.
This results in a single combined matrix which can then be used as input into edge density computations.
Statistical inference should then again be based on permutation testing over task labels
applied to edge density values.
We think that this approach would combine within-subject and between-subject variation in a logical way.
However, developing a statistical framework that fully implements statistical inference with multiple sources of variance
is beyond the scope of this paper, and will be the object of future work.
}

The interpretation of TED-results may not always be straightforward.
As can be seen from Fig.~\ref{diffsync}, synchronicity trajectories can take any shape, and may even show a downward slope following
stimulus onset. Therefore, strong hubs in TED hubness maps do not necessarily indicate strong activation. Also, as was the case
in the Gonzalez-Castillo study~\cite{GonzalezCastillo:2012js} - we now see many more brain regions that appear to be involved in the task.
The exact role of many of these regions and their mutual interactions are difficult to assess.
Furthermore, a single brain area found by TED may be involved in different task-dependent networks
and hence appear task-positive in reversed contrasts. We presented the right-hemi\-spheric motor area as an example
and show that this brain region
participates in different networks depending on the task (see Fig.~\ref{tedroi}). Note that a classical GLM analysis
is not able to depict such a scenario, even though such a reconfiguration is a highly
plausible rendition of human brain function. For example, in the case of fingertapping, we would expect to find
effects due to handedness so that the contrast ``left hand minus right hand'' should not simply be
the same as ``right hand minus left hand'' with the sign reversed.
And indeed, the TED hubness maps of the first contrast show a remarkably different pattern from that of the
reversed contrast -  an effect that may well be ascribed to handedness~\cite{handedness2007}.
In sum, the fact that we now have to deal with entire networks rather than univariate regions adds another level of
complexity to data interpretation and visualisation.
On the other hand, this complexity likely reflects much more closely the true complexity of human brain function.

\subsection*{Acknowledgments}
Data were provided by the Human Connectome Project, WU-Minn Consortium
(Principal Investigators: David Van Essen and Kamil Ugurbil; 1U54MH091657)
funded by the 16~NIH~Institutes and Centers that support the NIH~Blueprint for Neuroscience Research;
and by the McDonnell Center for Systems Neuroscience at Washington University.


\clearpage
\beginsupplement

\vspace*{4cm}

\section{Supplementary Information}

\clearpage

\begin{figure*}
\begin{center}
\includegraphics[width=1\textwidth]{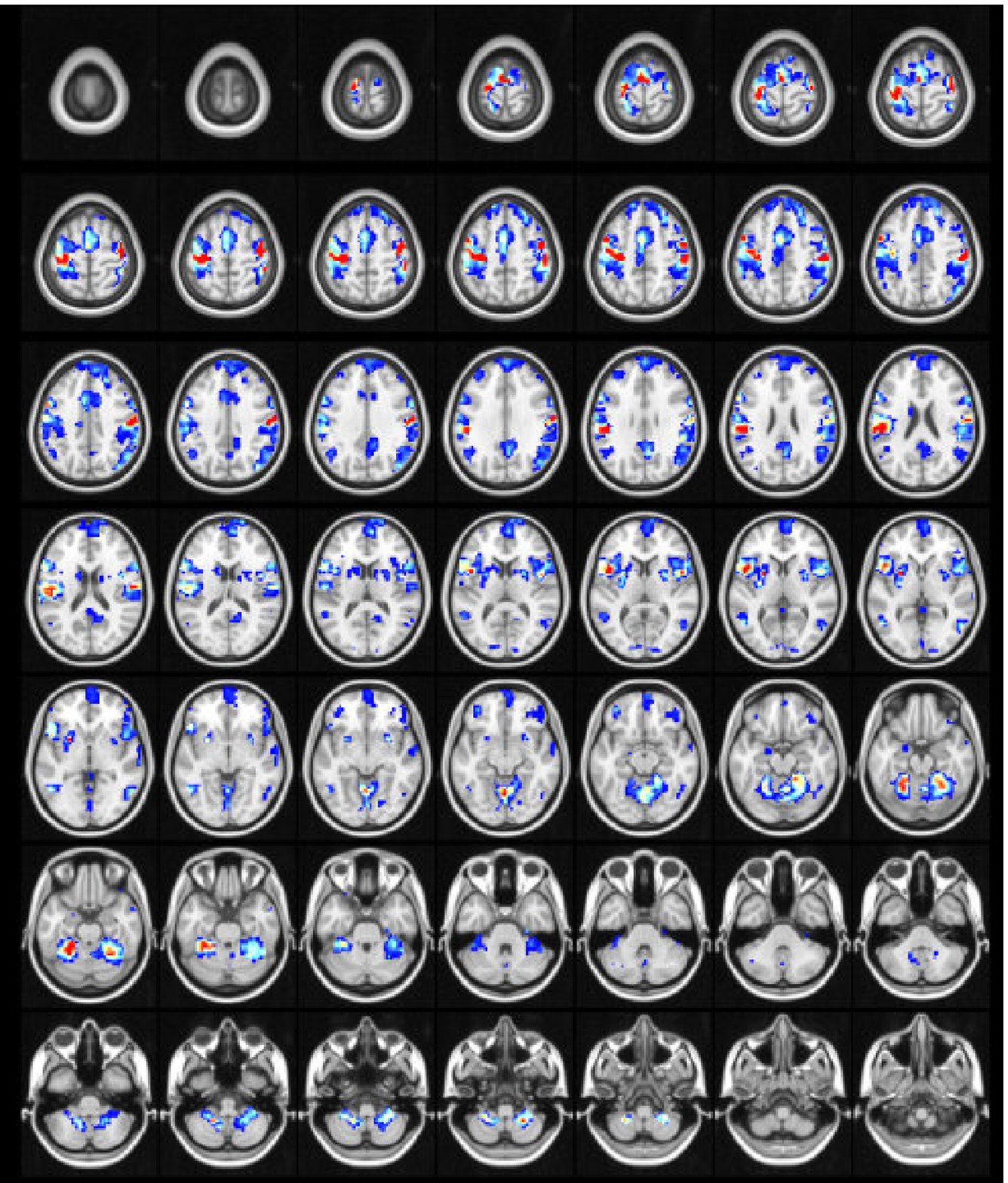}
\end{center}
\caption{{\bf Task-dependent dynamic reconfiguration of whole-brain networks}.
This map is based on the same data as figure~\ref{hubs_motor} of the main manuscript.
It shows the hubness map of the contrast left hand minus right hand fingertapping.}
\label{suppl_zlr}
\end{figure*}

\clearpage

\begin{figure*}
\begin{center}
\includegraphics[width=1\textwidth]{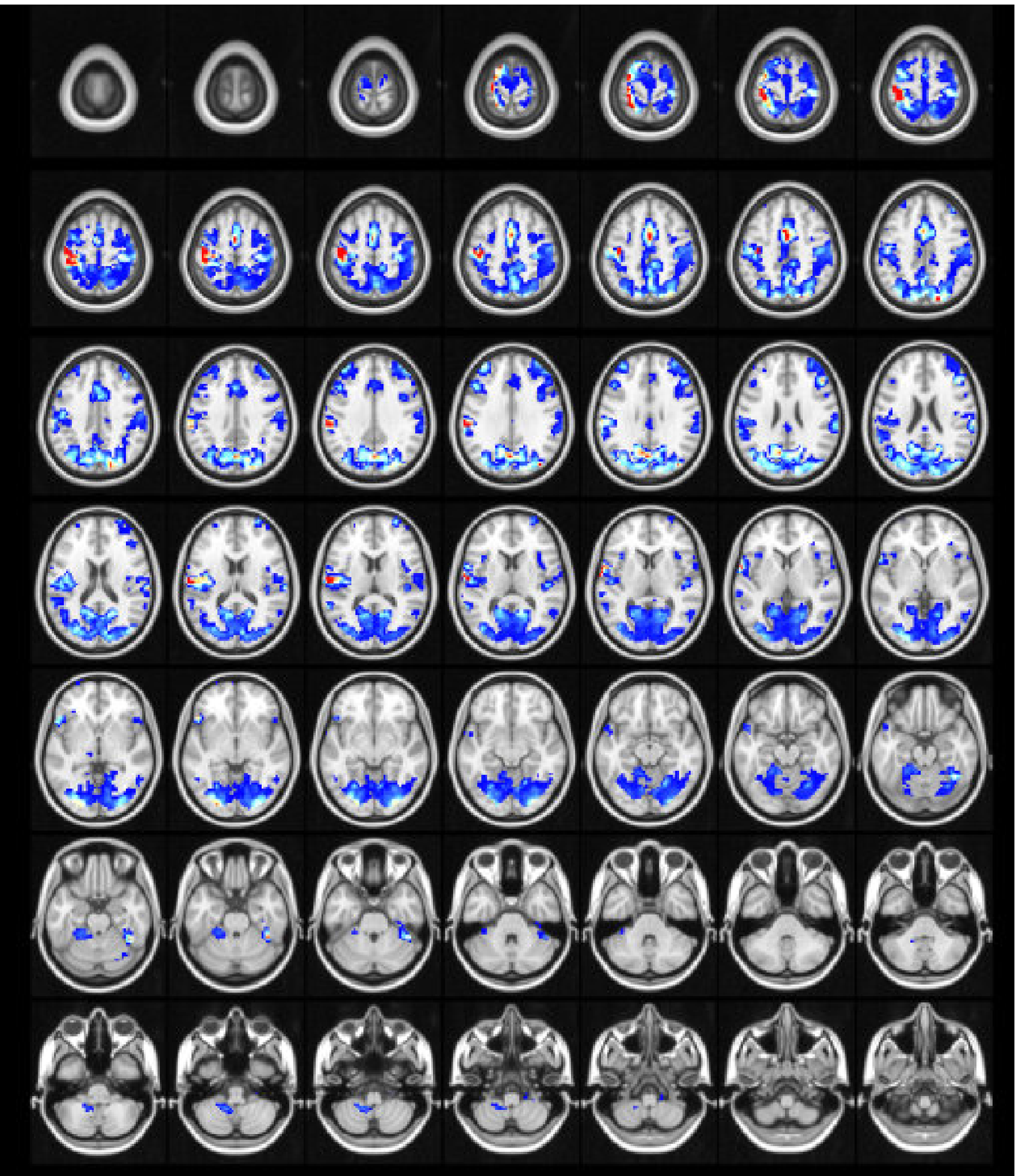}
\end{center}
\caption{{\bf Task-dependent dynamic reconfiguration of whole-brain networks}.
This map is based on the same data as figure~\ref{hubs_motor} of the main manuscript.
It shows the hubness map of the contrast right hand minus left hand fingertapping.}
\label{suppl_zrl}
\end{figure*}

\clearpage

\clearpage

\begin{figure*}
\begin{center}
\includegraphics[width=1\textwidth]{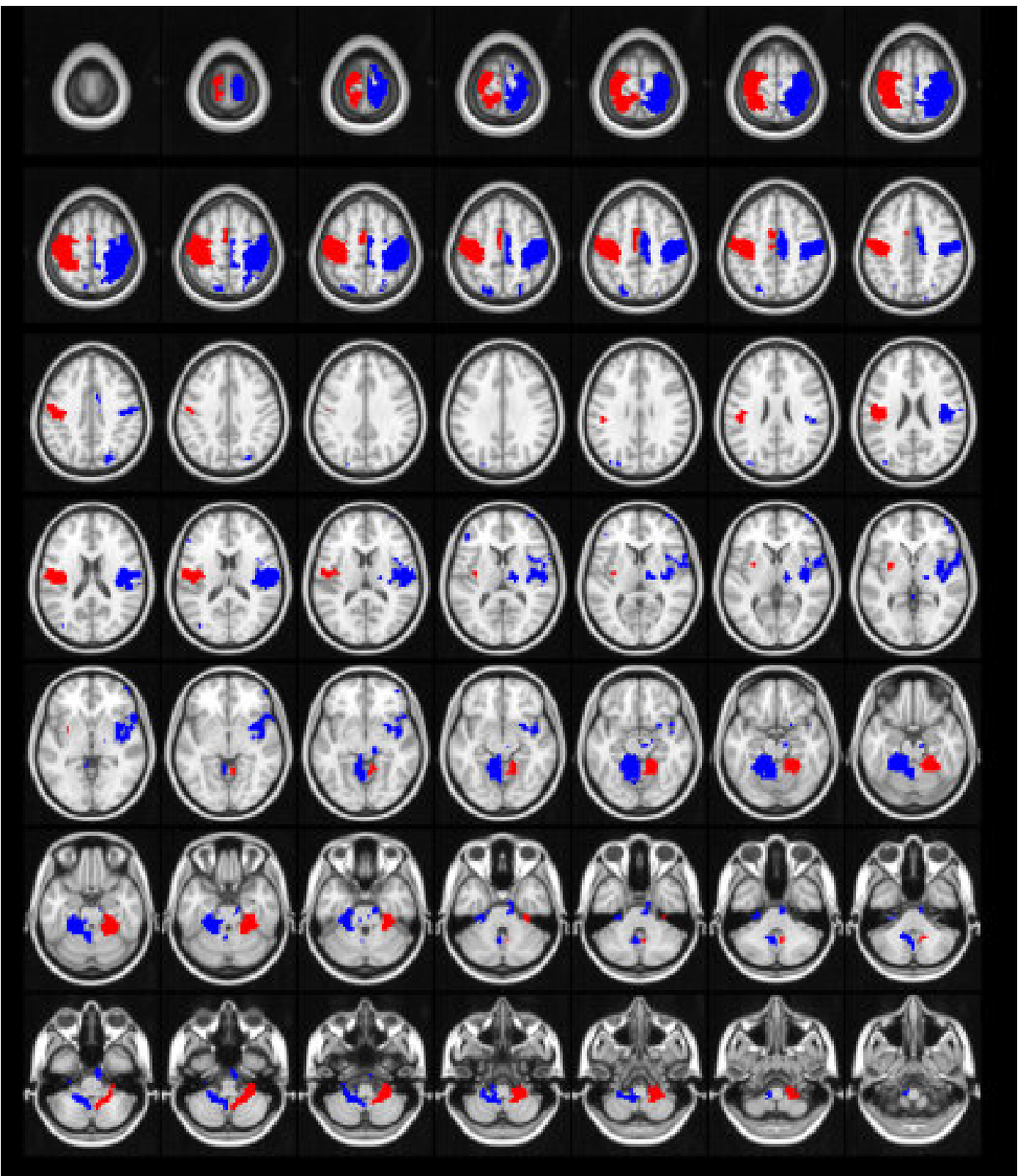}
\end{center}
\caption{{\bf GLM-based activation map.}
This map is based on the same data as figure~\ref{glm} of the main manuscript.}
\label{suppl_glm}
\end{figure*}

\end{document}